\documentclass[%
 reprint,
 amsmath,amssymb,
 aps,
pre,
floatfix,
]{revtex4-1}

\usepackage{graphicx}
\usepackage{dcolumn}
\usepackage{bm}

\usepackage{multirow}
\usepackage{array}

\DeclareMathOperator{\sech}{sech}

\begin{document}

\preprint{APS/123-QED}

\title{Replication of dissipative vortices modeled by the complex Ginzburg-Landau equation}

\author{Bogdan A. Kochetov$^1$}
\author{Vladimir R. Tuz$^{1,2}$}
\email{tvr@jlu.edu.cn; tvr@rian.kharkov.ua}
\affiliation{$^1$International Center of Future Science, State Key Laboratory on Integrated Optoelectronics, College of Electronic Science and Engineering, Jilin University, \\ 2699 Qianjin St., Changchun 130012, China}
\affiliation{$^2$Institute of Radio Astronomy of National Academy of Sciences of Ukraine, 4, Mystetstv St., Kharkiv 61002, Ukraine} 

\date{\today}

\begin{abstract}
Dissipative vortices are stable two-dimensional localized structures existing due to balance between gain and loss in nonlinear systems far from equilibrium. Being resistant to the dispersion and nonlinear distortions they are considered as promising information carriers for new optical systems. The key challenge in the development of such systems is getting control over vortex waveforms. In this paper we report on replication of two-dimensional fundamental dissipative solitons and vortices due to their scattering on a locally applied potential in the cubic-quintic complex Ginzburg-Landau equation. It has been found that an appropriate potential non-trivially splits both fundamental solitons and vortices into a few exact copies without losing in their amplitude levels. A remarkably simple potential having a finite supporter along the longitudinal coordinate and a double peaked dependence on a single transverse coordinate is found to be suitable for the replication of the two-dimensional localized structures.
\end{abstract}



\maketitle
\section{\label{intr}Introduction}
The complex Ginzburg-Landau equation (CGLE) implies a family of widespread models comprising gain, loss, nonlinearity, dispersion, and filtering -- all the fundamental features of nonlinear systems far from equilibrium that support the self-organization phenomena \cite{Cross_RMP_1993, Aranson_RMP_2002, García-Morales_CP_2012}. These models form the solid background for theoretical analysis of the structure formation in dissipative systems of diverse physical nature including active optical media \cite{Malomed_JOSAB_2014}, mode-locked lasers \cite{Grelu_NP_2012}, magneto-optic waveguides \cite{Boardman_1997, Boardman_Chapter_2005, Boardman_2006}, reaction-diffusion systems \cite{Liehr_Book}, Bose-Einstein condensates \cite{Malomed_Book, Ostrovskaya_PRA_2012, Smirnov_PRB_2014}, and others \cite{Akhmediev_Book1, Akhmediev_Book2, Purwins_AP_2010}.

Numerous studies of one-dimensional CGLE have continuously been aimed at searching new sophisticated solutions in the form of stable localized structures, which are now known as dissipative solitons \cite{Akhmediev_Chapter_2005}. Being totally dynamical objects due to dissipative processes, they may evolve as solitons with stationary \cite{Fauve_PRL_1990, van_Saarloos_PD_1992, Afanasjev_PRE_1996, Renninger_PRA_2008}, periodically, quasi-periodically, or aperiodically (chaotically) pulsating waveforms \cite{Deissler_PRL_1994, Soto-Crespo_PRL_2000, Akhmediev_PRE_2001}, exploding solitons \cite{Soto-Crespo_PRL_2000, Soto-Crespo_PLA_2001, Akhmediev_PRE_2001, Cundiff_PRL_2002, Descalzi_PRE_2011}, and solitons with periodical and chaotic spikes of extreme amplitude and short duration \cite{Chang_OL_2015, Chang_JOSAB_2015, Soto-Crespo_JOSAB_2017}. The one-dimensional cubic-quintic CGLE also admits multisoliton solutions \cite{Akhmediev_PRL_1997} and stable dynamic bound states of dissipative solitons \cite{Turaev_PRE_2007}. Many of these different types of dissipative solitons coexist in certain regions of the CGLE parameter space \cite{Afanasjev_PRE_1996, Soto-Crespo_PRL_2000, Soto-Crespo_PLA_2001, Akhmediev_PRE_2001}. Moreover, the complex dynamics of the CGLE supplemented with such higher-order terms as fourth-order spectral filtering, third-order dispersion, and stimulated Raman scattering has continuously attracted attention \cite{Soto-Crespo_PRE_2002, Achilleos_PRE_2016, Sakaguchi_OL_2018, Uzunov_PRE_2018}.

The two- and three-dimensional CGLEs have even greater variety of solutions manifesting nontrivial behaviors. First of all, it applies to stationary bullets, pulsating complexes, and dissipative vortices \cite{Crasovan_PRE_2000, Mihalache_PRL_2006, Soto-Crespo_OE_2006, Akhmediev_Chaos_2007, Mihalache_PRA_2007}. The latter ones are localized pulses whose phases have the form of rotating spirals with singularities. In optics, dissipative vortices are intriguing light beams with fascinating offers in applications \cite{Soskin_JO_2017}. Particularly, a dissipative optical vortex is considered as a promising information carrier, insensitive to the effects of dispersion and nonlinearity with the propagation distance and naturally quantized, due to a discrete set of allowable waveforms.

Apart from searching new types of solutions to the CGLEs, a large number of recent studies have been  targeted on investigation of various scenarios of complex dynamics, which result from interaction of dissipative solitons with external potentials. It has been accomplished in the framework of the CGLEs with added external potentials. For instance, using a CGLE model of a laser cavity, it has been revealed that localized vortices, built as sets of four peaks pinned to the periodic potential, may be stable without the diffusion term \cite{Leblond_PRA_2009}. In \cite{Sakaguchi_PRE_2009} a periodic potential has been incorporated into the CGLE to construct stable two-dimensional dissipative gap solitons in a bulk self-defocussing optical waveguide filled with a laser medium and equipped with transverse grating. The detailed analysis of interactions between moving dissipative solitons in multi-dimensional CGLEs with a linear potential has been carried out in \cite{He_OL_2009}. Different scenarios of the complex dynamics of one-dimensional dissipative solitons supported by a sharp potential barrier in the CGLE have been analyzed in \cite{He_JOSAB_2010}. Various dynamical regimes of the continuous generation of two-dimensional fundamental solitons in an active optical medium perturbed by the razor, dagger, and needle potentials in the CGLE have been presented in \cite{Liu_OL_2010}. The evolution of two-dimensional dissipative solitons on the top of externally applied umbrella-shaped and radial-azimuthal potentials in the CGLE has been analyzed in \cite{Yin_JOSAB_2011} and \cite{Liu_OE_2013}, respectively.

Furthermore, an external potential can break the time reversal symmetry as it takes place in planar magneto-optic waveguides \cite{Boardman_1997, Boardman_2001}, where an external magnetic field is applied to produce a potential, eventually leading to the non-reciprocal propagation of light beams. Indeed, the CGLE-based study of the influence of a spatially inhomogeneous external magnetic field upon propagation of dissipative solitons in a magneto-optic planar waveguide in the Voigt configuration has revealed significant benefits of utilizing the magnetic field to acquire different propagation conditions for the counter-propagating light beams \cite{Boardman_Chapter_2005, Boardman_2006}. Later on, this idea has been used to develop new robust control mechanisms for performing selective lateral shift within a group of stable noninteracting one-dimensional fundamental dissipative solitons \cite{OptLett_2017} as well as for their replication \cite{PRE_2017} and controllable transformations of soliton waveforms \cite{Chaos_2018}.

The success in the magneto-optic control over simple one-dimensional soliton waveforms raises our hopes to elaborate these ideas for the multi-dimensional case. First of all, it applies to gaining control over two-dimensional dissipative vortices in magneto-optical waveguide systems in the Faraday configuration. Controllable vortices insensitive to many distortion effects might become promising information carriers in the optical systems of the future. We pursue this goal in the present paper, and demonstrate a new mechanism for replication of both fundamental dissipative solitons and dissipative vortices assuming that their dynamics is governed by the two-dimensional cubic-quintic CGLE with a given potential. Particularly, this mechanism can be utilized in magneto-optic planar waveguides to design optical demultiplexers.

The rest of the paper is organized as follows. In Sec.~\ref{mod} we introduce the two-dimensional cubic-quintic CGLE with a potential term to employ it as a mathematical model allowing the control over waveforms of dissipative vortices. Sec.~\ref{sch} contains the description of an explicit computational scheme of the second order accuracy used in our numerical simulations. The replication of two-dimensional fundamental dissipative solitons caused by a locally applied potential is demonstrated in Sec.~\ref{rep1}. In Sec.~\ref{rep2} we present and discuss the peculiarities of replication of a dissipative vortex with unit topological charge and its corresponding antivortex. Conclusions and final remarks are summarized in Sec.~\ref{concl}.

\section{\label{mod}Model with Control}

We consider two-dimensional cubic-quintic CGLE with a potential, which can explicitly depend on all the coordinates. Having assumed that this model is used to describe an envelop of light beams in optical bulk media and nonlinear magneto-optic waveguide systems in the Faraday configuration, we adopt the notations used in \cite{He_JOSAB_2010, Liu_OL_2010} and \cite{Boardman_Chapter_2005, Boardman_2006}, respectively. Thus, the model under consideration is written in the following form
\begin{multline}
\label{CQCGLE}
\mathrm{i}\frac{\partial\Psi}{\partial z}+\mathrm{i}\delta\Psi+\left(\frac{D}{2}-\mathrm{i}\beta\right)\left(\frac{\partial^2\Psi}{\partial x^2}+\frac{\partial^2\Psi}{\partial y^2}\right)
+\left(1-\mathrm{i}\varepsilon\right)\left|\Psi\right|^2\Psi\\ - \left(\nu-\mathrm{i}\mu\right)\left|\Psi\right|^4\Psi + Q(x,y,z)\Psi=0, 
\end{multline}
where $\Psi\left(x,y,z\right)$ is an unknown complex amplitude of the transverse $x$, $y$ and the longitudinal $z$ coordinates, $D$ is the group velocity dispersion coefficient, $\delta$ is the linear absorption coefficient, $\beta$ is the linear diffusion coefficient, $\varepsilon$ is the coefficient of nonlinear cubic gain, $\nu$ accounts for the self-defocusing effect, and $\mu$ defines quintic nonlinear losses.

Generally, the potential $Q(x,y,z)$ in the model~(\ref{CQCGLE}) can be used to account for an arbitrary linear control over soliton waveforms. Depending on the physical origin of the problem under consideration the potential acquires the required profile. In this regard we can mention the magnetization function in magneto-optic systems \cite{Boardman_1997, Boardman_Chapter_2005, Boardman_2006, OptLett_2017, PRE_2017, Chaos_2018}, inhomogeneity of the refractive index in bulk optical media and lattices \cite{Holmer_JNS_2007, Yang_OE_2008, He_OL_2009, Leblond_PRA_2009, Sakaguchi_PRE_2009, Liu_OL_2010, He_JOSAB_2010, Yin_JOSAB_2011, Liu_OE_2011, Liu_OE_2013}, and potentials in Bose-Einstein condensates \cite{Malomed_Book, Ostrovskaya_PRA_2012, Smirnov_PRB_2014}.

\begin{figure}[t]
\centering
\includegraphics[width=\linewidth]{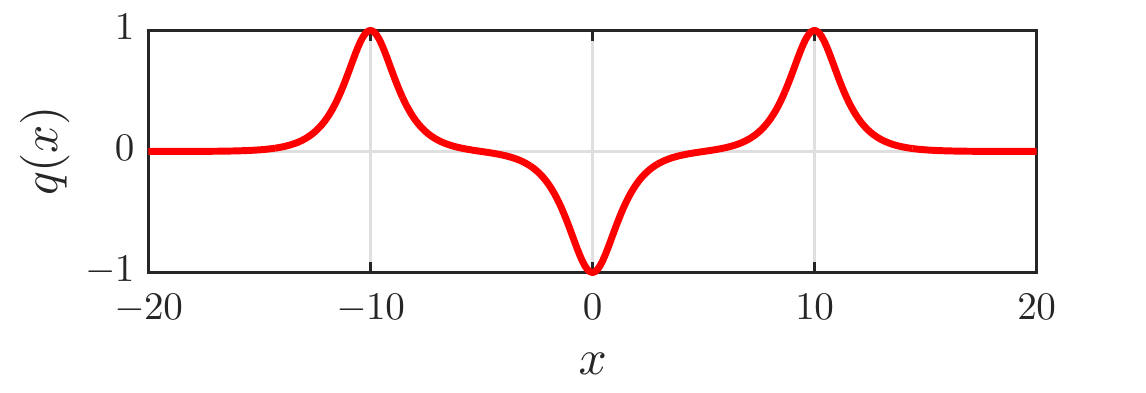}
\caption{Transverse dependence of the potential for $x_p=10$.}
\label{fig1}
\end{figure}

In order to choose a particular profile we assume that the potential $Q$ corresponds to an externally induced magnetization in a magneto-optic system in the Faraday configuration. Here we consider the case when the potential does not explicitly depend on the $y$ coordinate and it is applied locally (it has a finite supporter) along the $z$ coordinate. For simplicity, having chosen its longitudinal dependence in the form of a piecewise constant function we write down
\begin{equation}
\label{Q}
Q(x,z) = q(x)\left[h(z-z_1)-h(z-z_2)\right],
\end{equation}
where $q(x)$ describes the potential dependence on the transverse $x$ coordinate, $h(\cdot)$ is the Heaviside step function, while $z_1$ and $z_2$ stand for the two points on the $z$ axis at which the potential is switched on and switched off, respectively. We choose an inhomogeneous spatial distribution of the magnetization in a magneto-optic system in the form of a few potential wells arranged along the $x$ axis. The shape of each potential well is expressed by the $\sech(x)$ function, which is a suitable trial function approximating the experimental distributions of the magnetic fields within magneto-optic systems \cite{Boardman_1997, Boardman_Chapter_2005, Boardman_2006}. We take the $q(x)$ function in the following form to get a symmetric repulsive potential in the vicinity of $x=0$ point
\begin{equation}
\label{q}
q(x) = \sech(x-x_p)-\sech(x)+\sech(x+x_p),
\end{equation}
where $\pm x_p$ are the $x$ coordinates of two potential peaks, wheres the single dip is located at the $x=0$ point. Thus, an inhomogeneous spatial distribution of the potential $Q$ has the form of $x$-arranged potential wells whose depths are non-zero constants within the interval $z\in[z_1,z_2]$. A particular profile of the $x$-dependent potential barrier \eqref{q} for $x_p=10$ is presented in Fig.~\ref{fig1}.

Assuming the periodicity of the complex amplitude $\Psi$ in both transverse directions we supplement Eq.~\eqref{CQCGLE} with the following periodic boundary conditions 
\begin{equation}
\label{PBC}
\Psi(x,y,z)=\Psi(x+L_x,y,z)=\Psi(x,y+L_y,z),
\end{equation}
$\forall~(x,y,z)\in\mathbb{R}^2\times\left[0,L_z\right]$, for some $L_x$, $L_y$, $L_z>0$.

In order to excite a stable dissipative vortex in the system \eqref{CQCGLE} any function whose shape is sufficiently close to the vortex waveform can be taken as an initial condition. Thus, in all our numerical simulations we use the following initial condition
\begin{equation}
\label{IC}
\Psi(x,y,0)=A_0\left[x+\mathrm{i}y\frac{m}{|m|}\right]^{|m|}\exp\left(-\frac{x^2+y^2}{r_0^2}\right),
\end{equation}
where $A_0$ and $r_0$ adjust the height and width of the initial waveform, respectively, whereas $m$ defines the topological charge of a vortex, which is the vortex key property defining the winding number of a wavefront around its core.

\section{\label{sch}Numerical Scheme}
We solve the problem \eqref{CQCGLE}-\eqref{IC} combining the fast Fourier transform for inversion of the linear terms in Eq.~\eqref{CQCGLE} and the exponential time differencing (ETD) method \cite{Cox_2002} to solve the corresponding nonlinear ODE with the potential term in the Fourier domain. Previously, we applied this approach in the one-dimensional case \cite{Chaos_2018}.

Since each solution to Eq.~\eqref{CQCGLE} satisfies to the periodic boundary conditions \eqref{PBC}, the computational domain is reduced to the form $\left[-L_x/2,L_x/2\right)\times\left[-L_y/2,L_y/2\right)\times\left[0,L_z\right]$. The transverse sizes of computational domain we choose to ensure that any non-negligible part of soliton waveforms is located completely within the domain. In order to compute the two-dimensional fast Fourier transform we sample the intervals along the $x$ and $y$ coordinates with $N_x=N_y=2^9$ discretization points, while the distance along the $z$ coordinate is discretized with the step $\Delta z=0.005$. 

The second-order ETD scheme for updating the Fourier image of complex amplitude on the grid along the $z$ axis has the form
\begin{equation}
\label{ETD2}
\begin{split}
\hat{\Psi}_{n+1}=\hat{\Psi}_{n}e^{\sigma\Delta z}+\hat{\mathcal{N}}_{n-1}\dfrac{1+\sigma\Delta z-e^{\sigma\Delta z}}{\sigma^2\Delta z}\\
+\hat{\mathcal{N}}_n\dfrac{\left(1+\sigma\Delta z\right)e^{\sigma\Delta z}-1-2\sigma\Delta z}{\sigma^2\Delta z},\\
\mathcal{N}(\Psi,x,y,z)\\
=\left[\left(\varepsilon+\mathrm{i}\right)\left|\Psi\right|^2
-\left(\mu+\mathrm{i}\nu\right)\left|\Psi\right|^4+\mathrm{i}Q(x,y,z)\right]\Psi,
\end{split}
\end{equation}
where $\sigma=-\delta-\left(\beta+\mathrm{i}D/2\right)(k_x^2+k_y^2)$ is a spectral parameter, $\Psi_n=\Psi(x,y,z_n)$, $\mathcal{N}_n=\mathcal{N}\left(\Psi_n,x,y,z_n\right)$, $z_n=n\Delta z$, and the circumflex denotes the two-dimensional discrete Fourier transform with respect to both transverse coordinates, i.e. $\hat{\Psi}(k_x,k_y,z)= \mathcal{F}\left\lbrace\Psi(x,y,z)\right\rbrace$, $\hat{\mathcal{N}}(k_x,k_y,z)= \mathcal{F}\left\lbrace \mathcal{N}(\Psi(x,y,z),x,y,z)\right\rbrace$.

In all our numerical simulations presented here, we fix and always use the same values for the following parameters of Eq.~\eqref{CQCGLE}, the potential \eqref{Q}, \eqref{q}, the initial condition \eqref{IC}, and the computational domain, which are summarized in Table~\ref{tab1}. These chosen parameters of Eq.~\eqref{CQCGLE} and the initial waveform \eqref{IC} admit the excitation and fast development of both fundamental solitons and vortices in the model \eqref{CQCGLE}-\eqref{IC}. Also we chose them taking into account the previous studies, where the same set of the equation parameters has been analyzed \cite{Afanasjev_PRE_1996} and used for modeling \cite{Boardman_Chapter_2005}.
\begin{table}[htbp]
\caption{Fixed parameters used in numerical simulations.}
\label{tab1}
\centering
\begin{tabular}{|c|c|c|c|c|c|}
\hline
\multicolumn{6}{|c|}{\textbf{Parameters of Equation}}\\
\hline
$~D=1~$ & $~\beta=0.5~$ & $~\delta=0.5~$ & $~\mu=1~$ & $~\nu=0.1~$ & $~\varepsilon=2.5~$\\
\hline
\hline
\multicolumn{6}{|c|}{\textbf{Parameters of Potential}}\\
\hline
\multicolumn{3}{|c|}{$x_p=10$} & \multicolumn{3}{|c|}{$z_1=50$}\\
\hline
\hline
\multicolumn{6}{|c|}{\textbf{Parameters of Initial Waveform}}\\
\hline
\multicolumn{3}{|c|}{$A_0=1$} & \multicolumn{3}{|c|}{$r_0=3$}\\
\hline
\hline
\multicolumn{6}{|c|}{\textbf{Computational Domain}}\\
\hline
\multicolumn{2}{|c|}{$L_x=60$} & \multicolumn{2}{|c|}{$L_y=60$} & \multicolumn{2}{|c|}{$L_z=200$}\\
\hline
\multicolumn{2}{|c|}{$N_x=512$} & \multicolumn{2}{|c|}{$N_y=512$} & \multicolumn{2}{|c|}{$\Delta z=0.005$}\\
\hline
\end{tabular}
\end{table}

On the other hand, the vortex topological charge $m$ in Eq.~\eqref{IC} and the longitudinal coordinate $z_2$ (it stands for the parameter controlling the replication) at which the potential~\eqref{Q} is switched off are different for each simulation.

\section{\label{rep}Replication of 2D Dissipative Solitons}
In this section, we demonstrate new results for splitting two-dimensional localized structures supported by the model \eqref{CQCGLE}-\eqref{IC} into a few replicas. Namely, we are focused on the study of replication of two-dimensional fundamental dissipative solitons as well as dissipative vortices with the topological charges $m=\pm 1$ employing the numerical scheme~\eqref{ETD2}. We subsequently consider these two cases in Secs.~\ref{rep1} and \ref{rep2}, respectively. 

\subsection{\label{rep1}Replication of Fundamental Solitons}

Here we consider a local influence of an external potential upon dynamics of two-dimensional fundamental dissipative solitons $(m=0)$ to find and demonstrate a nontrivial scenario of their evolution leading to the soliton replication -- splitting of a single soliton into a few exact copies.

\begin{figure}[ht]
\centering
\includegraphics[width=\linewidth]{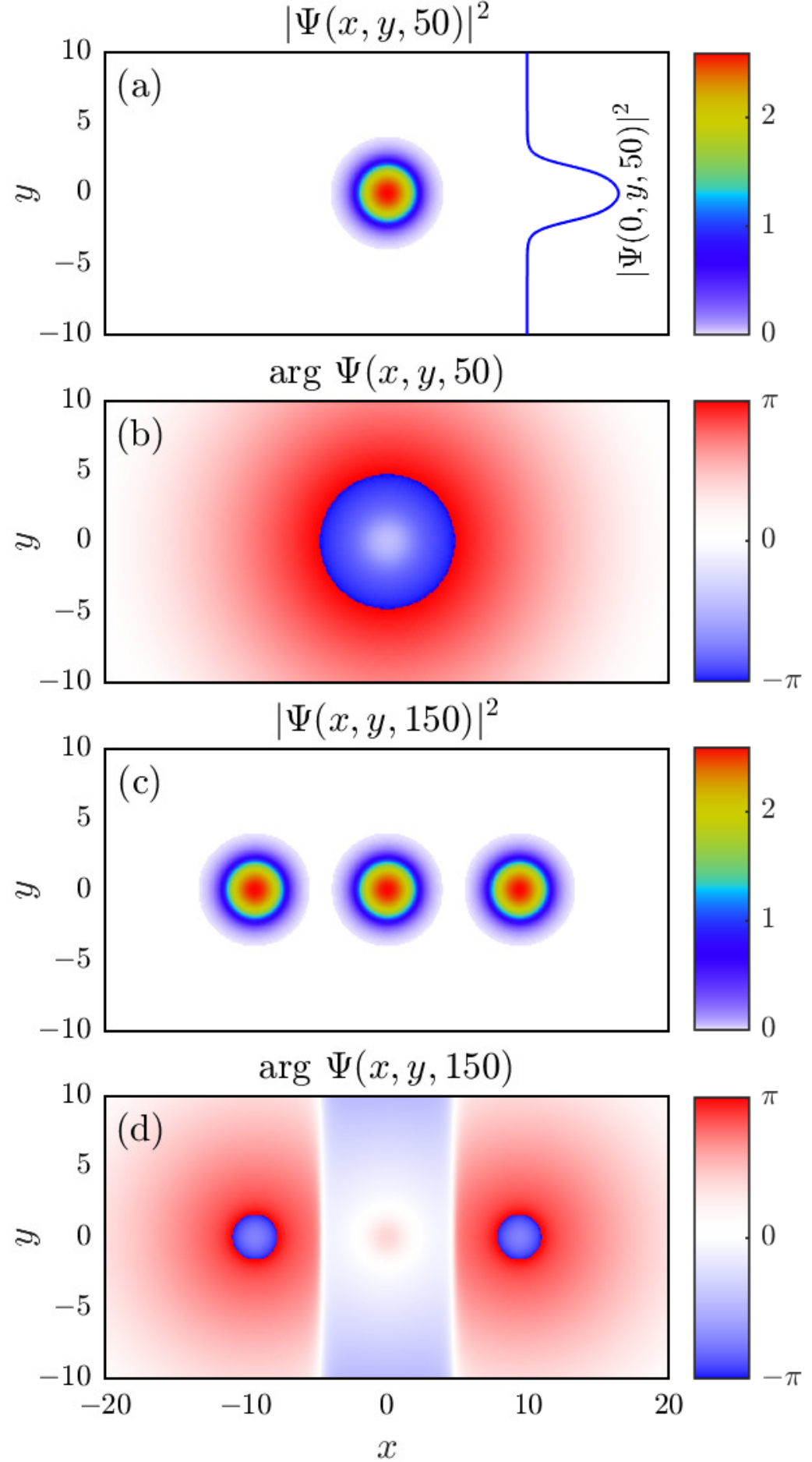}
\caption{Replication of a two-dimensional fundamental dissipative soliton $(m=0)$ due the local influence of the potential applied along the $z$ axis between two points $z_1=50$ and $z_2=83$; (a) intensity plot and (b) phase diagram of a seed soliton at $z=50$, where the potential is switched on; (c) intensity plot and (d) phase diagram of three replicated copies at $z=150$, where perturbations induced by the potential have decayed. The insert in panel (a) shows the cross section of the intensity plot at $y=0$.}
\label{fig2}
\end{figure}

In particular, for the parameters listed in Table~\ref{tab1}, the replication of a fundamental soliton into three exact copies due to its interaction with the potential \eqref{Q}, \eqref{q} has been found when the potential is switched off in a vicinity of the point $z_2=83$. This effect is illustrated in Fig.~\ref{fig2}. The intensity plots of the squared absolute values of the complex amplitudes of both single seed soliton and its three replicated copies are plotted in Figs.~\ref{fig2}(a) and \ref{fig2}(c), respectively. The phase diagrams of the corresponding complex amplitudes are presented in Figs.~\ref{fig2}(b) and \ref{fig2}(d). Additionally, the animation showing the dynamics of this replication of the fundamental soliton into three exact copies can be found in the Supplemental Material \cite{Suppl_Mat}.

The soliton replication has been performed in three stages. First, the initial waveform \eqref{IC} with $m=0$ arises at the section $z=0$, and then quickly evolves to the fundamental soliton by the section $z=z_1=50$, where the potential \eqref{Q}, \eqref{q} comes into effect. We note that the approaching of the initial waveform to an attractor is asymptotic, and the development of an exact fundamental soliton is only possible in an infinite time. However, in all the numerical simulations the solution $\Psi(x,y,50)$ presented in Figs.~\ref{fig2}(a) and \ref{fig2}(b) is indistinguishable from the exact soliton due to a finite accuracy of the numerical method. Therefore, we identify this numerical solution as the developed fundamental soliton. Second, between the two sections $z=z_1=50$ and $z=z_2=83$ the applied potential perturbs the fundamental soliton and changes its waveform dramatically. Third, at the section $z=z_2=83$ the potential $Q$ is switched off, and the perturbed waveform rapidly transits to a new stable state in the form of three noninteracting fundamental dissipative solitons, which have identical waveforms as had the seed (unperturbed) soliton before the potential switched on. The intensity plot and phase diagram of the solution containing these three solitons are calculated at the section $z=150$ and plotted in Figs.~\ref{fig2}(c) and \ref{fig2}(d), respectively. 

The replication of dissipative solitons can also be viewed through the theory of dynamical systems. From this viewpoint, a fundamental dissipative soliton admitted by Eq.~\eqref{CQCGLE} is identified as an attractor that exists in the infinite-dimensional phase space of the system \eqref{CQCGLE}. For a fixed set of equation parameters, each attractor has certain set of initial conditions forming its basin of attraction. For instance, the initial waveform \eqref{IC} with $m=0$ evolves to the fundamental solution with spatial waveform distribution as presented in Fig.~\ref{fig2}(a) because the initial waveform belongs to the soliton's basin of attraction and never goes beyond it as long as the system parameters are fixed. On the other hand, the influence of a potential upon the soliton dynamics can be strong enough, leading to the transition of soliton waveform between basins of attraction of different attractors coexisting under the same set of system parameters as well as spatial splitting of the waveform into a few ones suitable for the emergence and further development of new noninteracting solitons. In other words, at the point at which the potential is switched off the perturbed soliton waveform can be considered as a new initial waveform whose further evolution is defined by a basin of attraction that contains it. The former scenario can be used to perform nontrivial transitions between soliton waveforms accompanied by the change of topological charge. For example, it is a transformation of a dissipative vortex with some charge to a vortex with another charge. The latter scenario allows to control both the multiple replication of a seed soliton and generation of solitons of new types. 

The replication of fundamental soliton presented in Fig.~\ref{fig2} is performed according to the latter scenario, because in the vicinity of the section $z=83$ the perturbed waveform is split into three spaced bell-shaped profiles such that each of them simultaneously belongs to the same basin of attraction as the initial waveform \eqref{IC}. Therefore, having switched off the potential at the section $z=83$ one gets a practical example of triple replication of the fundamental soliton.

Recently, similar effects related to both scenarios of evolution of one-dimensional dissipative solitons have been studied in \cite{PRE_2017, Chaos_2018}. The example presented in Fig.~\ref{fig2} confirms the existence of the same scenario for the replication in the case of two-dimensional fundamental solitons.

\subsection{\label{rep2}Replication of Vortices}
Having found the replication of fundamental dissipative solitons we make the next logical step: searching for the replication of dissipative vortices. Namely, we consider the case when a vortex has the topological charge of unit magnitude, i.e. $m=\pm 1$.

\begin{figure}[ht]
\centering
\includegraphics[width=\linewidth]{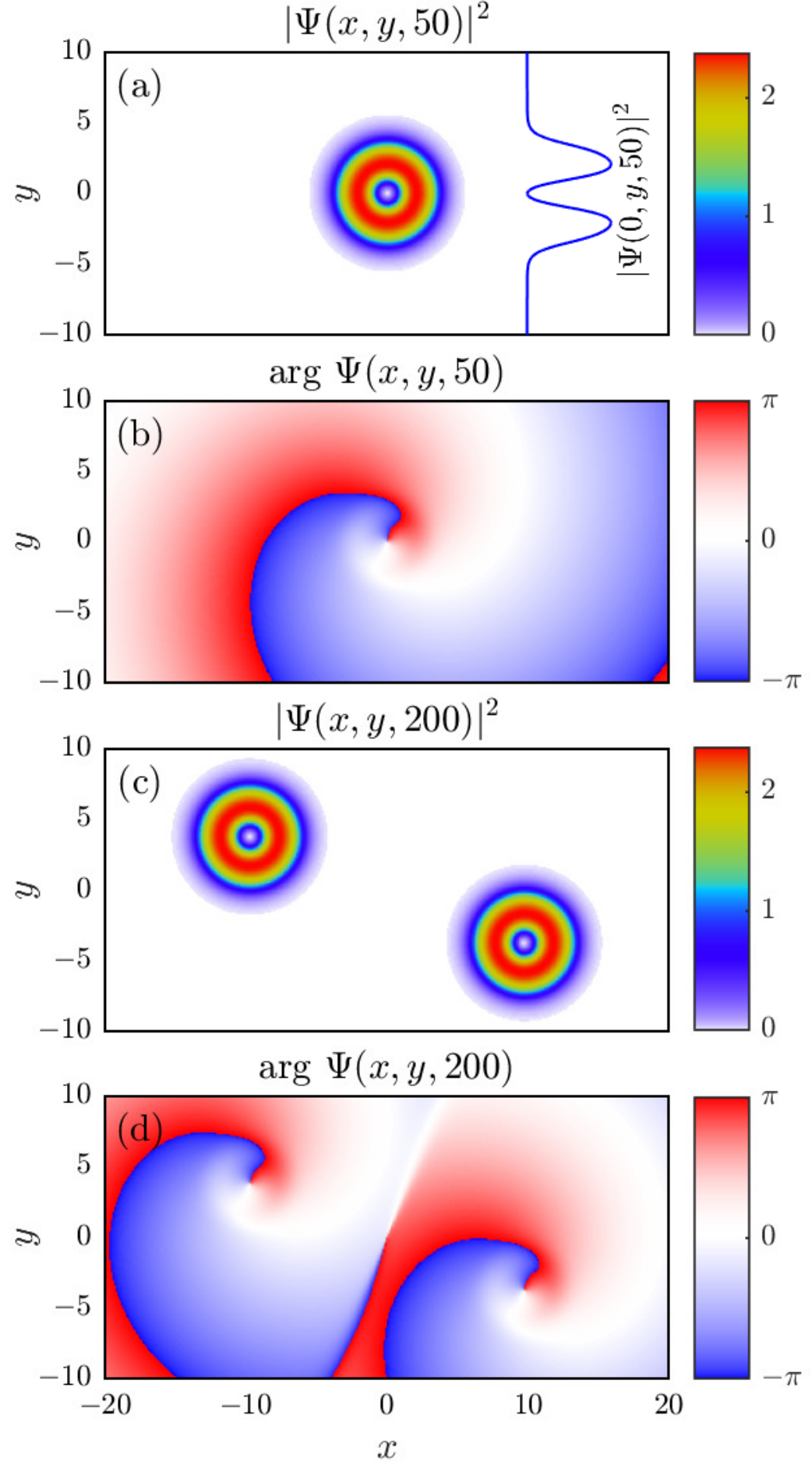}
\caption{Replication of a dissipative vortex with the topological charge $m=1$ due the local influence of the potential applied along the $z$ axis between two points $z_1=50$ and $z_2=133$; (a) intensity plot and (b) phase diagram of a single seed vortex at $z=50$, where the potential is switched on; (c) intensity plot and (d) phase diagram of two replicated copies at $z=200$, where perturbations induced by the potential have decayed. The insert in panel (a) indicates the cross section of the intensity plot at $y=0$.}
\label{fig3}
\end{figure}

\begin{figure}[ht]
\centering
\includegraphics[width=\linewidth]{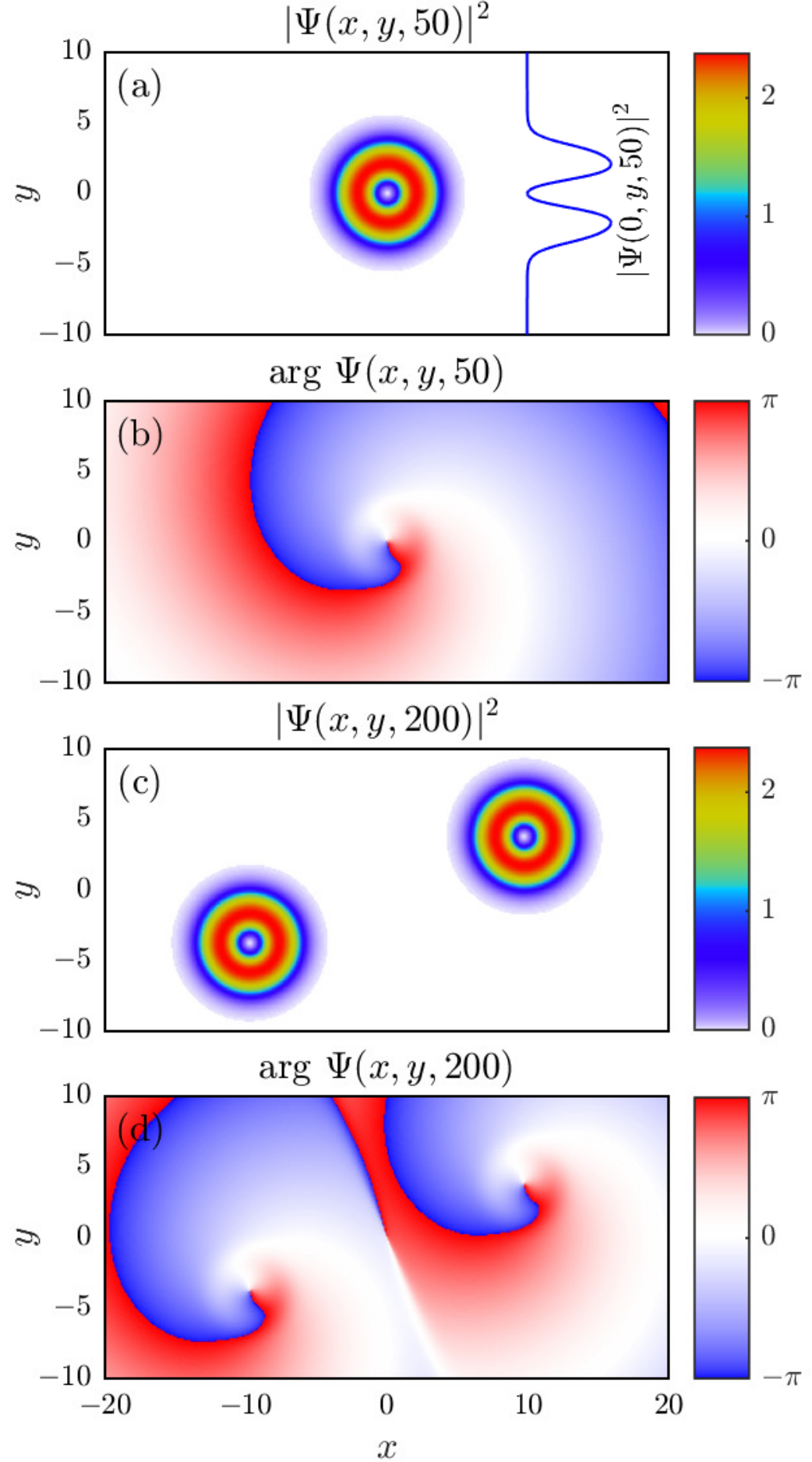}
\caption{The same as in Fig.~\ref{fig3} only for a dissipative antivortex with the topological charge $m=-1$.}
\label{fig4}
\end{figure}

Remarkably, applying the potential \eqref{Q} with the function \eqref{q} we can as well replicate a dissipative vortex with $m=1$ into two exact copies if the potential is switched off at the point $z_2=133$ as demonstrated in Fig.~\ref{fig3} and Supplemental Material \cite{Suppl_Mat}. Moreover, the same potential ($z_2=133$) can be applied to replicate the corresponding antivortex, i.e. the vortex with $m=-1$ (see Fig.~\ref{fig4} and Supplemental Material \cite{Suppl_Mat}). The plots in Fig.~\ref{fig3} and Fig.~\ref{fig4} are similar to those illustrated in Fig.~\ref{fig2}. The intensity plot and the phase diagram of the seed vortex (antivortex) calculated at the section $z=z_1=50$, at which the potential is switched on, are presented in Figs.~\ref{fig3}(a) and \ref{fig3}(b) (Figs.~\ref{fig4}(a) and \ref{fig4}(b)), respectively. On the other hand, Figs.~\ref{fig3}(c) and \ref{fig3}(d) (Figs.~\ref{fig4}(c) and \ref{fig4}(d)) show respectively the intensity plot and phase diagram of two replicated copies of the vortex (antivortex) calculated at the section $z=200$, at which both replicas correspond to the developed vortices (antivortices). 

The replications of fundamental soliton (Fig.~\ref{fig2}), vortex (Fig.~\ref{fig3}) and antivortex (Fig.~\ref{fig4}) have been performed according to the same scenario described in Sec.~\ref{rep1}. In all cases the key role is played by localized repulsive inhomogeneity of the potential along the transverse $x$ coordinate and the potential locality along the propagation distance $z$. The inhomogeneity along the $x$ axis is necessary to induce a nonuniform perturbation of soliton waveform. For sufficiently strong perturbations, the spatial waveform distribution can evolve to such profile that instantaneously is an appropriate initial condition to excite a few noninteracting solitons in the absence of an external potential in the system \eqref{CQCGLE}. Such profiles are evolved in the vicinities of points $z=83$ and $z=133$ leading to the development of three fundamental solitons and two vortices (antivortices) from the perturbed waveforms, when the potential is switched off at these points. The potential locality along the $z$ axis is needed to remove the potential influence upon solitons as soon as the desired perturbation of soliton waveform is achieved.

We note that the development of perturbed waveform distributions leading to splitting of dissipative solitons is a nontrivial dynamical process, which strongly depends on the applied potential. For example, the replication of fundamental solitons presented in Fig.~\ref{fig2} can be performed if the potential is only switched off within zones, quasi-periodically located along the $z$ axis. Those zones contain the waveforms suitable for the triple replication of the fundamental soliton along the propagation distance. Recently the existence of similar zones supporting the replication of one-dimensional solitons has been demonstrated in \cite{PRE_2017, Chaos_2018}. However, the replication of the dissipative vortex (Fig.~\ref{fig3}) and antivortex (Fig.~\ref{fig4}) is only possible if the potential \eqref{Q}, \eqref{q} is switched off in vicinities of a few isolated points, one of them being $z=133$. Thus, the replication of dissipative solitons and vortices have a common property of being very sensitive to potential profile change.

\section{\label{concl}Conclusions}
We considered the effect of replication of both dissipative vortices with a unit in magnitude topological charge ($m=\pm 1$) and fundamental solitons ($m=0$) in a system modeled by the two-dimensional cubic-quintic CGLE with a potential term supplemented to gain control over soliton waveform evolution. A potential suitable for replication should satisfy the two following requirements. First, the potential should have an inhomogeneous repulsive dependence on the transverse coordinates to transform a given soliton waveform into an initial condition appropriate for the development of a few noninteracting copies of the seed soliton when there is no applied potential in the system. Second, the potential should be applied locally along the longitudinal coordinate $z$ to stop its influence upon a soliton waveform as soon as the appropriate initial condition is developed.

From our numerous numeric simulations we can conclude that the replication of two-dimensional dissipative solitons can occur both in vicinities of a few isolated points and in periodically (quasi-periodically) located zones. However, we could not find the potential profile, for which a perturbed soliton waveform transits to a new state, which would admit the replication independently of the longitudinal coordinate $z$ at which the potential is switched off. Thus, the question about existence of such potential profiles is still open both for one- and multi-dimensional CGLEs.

We analyzed the replication of dissipative vortices in the framework of the particular two-dimensional CGLE comprising the basic properties of any dissipative system. Therefore, the replication of two-dimensional localized structures is logically expected to be found in other models manifesting a common dissipative behavior, beyond the CGLE.

\bibliography{Vortex}

\begin{thebibliography}{57}%
\makeatletter
\providecommand \@ifxundefined [1]{%
 \@ifx{#1\undefined}
}%
\providecommand \@ifnum [1]{%
 \ifnum #1\expandafter \@firstoftwo
 \else \expandafter \@secondoftwo
 \fi
}%
\providecommand \@ifx [1]{%
 \ifx #1\expandafter \@firstoftwo
 \else \expandafter \@secondoftwo
 \fi
}%
\providecommand \natexlab [1]{#1}%
\providecommand \enquote  [1]{``#1''}%
\providecommand \bibnamefont  [1]{#1}%
\providecommand \bibfnamefont [1]{#1}%
\providecommand \citenamefont [1]{#1}%
\providecommand \href@noop [0]{\@secondoftwo}%
\providecommand \href [0]{\begingroup \@sanitize@url \@href}%
\providecommand \@href[1]{\@@startlink{#1}\@@href}%
\providecommand \@@href[1]{\endgroup#1\@@endlink}%
\providecommand \@sanitize@url [0]{\catcode `\\12\catcode `\$12\catcode
  `\&12\catcode `\#12\catcode `\^12\catcode `\_12\catcode `\%12\relax}%
\providecommand \@@startlink[1]{}%
\providecommand \@@endlink[0]{}%
\providecommand \url  [0]{\begingroup\@sanitize@url \@url }%
\providecommand \@url [1]{\endgroup\@href {#1}{\urlprefix }}%
\providecommand \urlprefix  [0]{URL }%
\providecommand \Eprint [0]{\href }%
\providecommand \doibase [0]{http://dx.doi.org/}%
\providecommand \selectlanguage [0]{\@gobble}%
\providecommand \bibinfo  [0]{\@secondoftwo}%
\providecommand \bibfield  [0]{\@secondoftwo}%
\providecommand \translation [1]{[#1]}%
\providecommand \BibitemOpen [0]{}%
\providecommand \bibitemStop [0]{}%
\providecommand \bibitemNoStop [0]{.\EOS\space}%
\providecommand \EOS [0]{\spacefactor3000\relax}%
\providecommand \BibitemShut  [1]{\csname bibitem#1\endcsname}%
\let\auto@bib@innerbib\@empty
\bibitem [{\citenamefont {Cross}\ and\ \citenamefont
  {Hohenberg}(1993)}]{Cross_RMP_1993}%
  \BibitemOpen
  \bibfield  {author} {\bibinfo {author} {\bibfnamefont {M.~C.}\ \bibnamefont
  {Cross}}\ and\ \bibinfo {author} {\bibfnamefont {P.~C.}\ \bibnamefont
  {Hohenberg}},\ }\href {\doibase 10.1103/RevModPhys.65.851} {\bibfield
  {journal} {\bibinfo  {journal} {Rev. Mod. Phys.}\ }\textbf {\bibinfo {volume}
  {65}},\ \bibinfo {pages} {851} (\bibinfo {year} {1993})}\BibitemShut
  {NoStop}%
\bibitem [{\citenamefont {Aranson}\ and\ \citenamefont
  {Kramer}(2002)}]{Aranson_RMP_2002}%
  \BibitemOpen
  \bibfield  {author} {\bibinfo {author} {\bibfnamefont {I.~S.}\ \bibnamefont
  {Aranson}}\ and\ \bibinfo {author} {\bibfnamefont {L.}~\bibnamefont
  {Kramer}},\ }\href {\doibase 10.1103/RevModPhys.74.99} {\bibfield  {journal}
  {\bibinfo  {journal} {Rev. Mod. Phys.}\ }\textbf {\bibinfo {volume} {74}},\
  \bibinfo {pages} {99} (\bibinfo {year} {2002})}\BibitemShut {NoStop}%
\bibitem [{\citenamefont {García-Morales}\ and\ \citenamefont
  {Krischer}(2012)}]{García-Morales_CP_2012}%
  \BibitemOpen
  \bibfield  {author} {\bibinfo {author} {\bibfnamefont {V.}~\bibnamefont
  {García-Morales}}\ and\ \bibinfo {author} {\bibfnamefont {K.}~\bibnamefont
  {Krischer}},\ }\href@noop {} {\bibfield  {journal} {\bibinfo  {journal}
  {Contemporary Physics}\ }\textbf {\bibinfo {volume} {53}},\ \bibinfo {pages}
  {79} (\bibinfo {year} {2012})}\BibitemShut {NoStop}%
\bibitem [{\citenamefont {Malomed}(2014)}]{Malomed_JOSAB_2014}%
  \BibitemOpen
  \bibfield  {author} {\bibinfo {author} {\bibfnamefont {B.~A.}\ \bibnamefont
  {Malomed}},\ }\href {\doibase 10.1364/JOSAB.31.002460} {\bibfield  {journal}
  {\bibinfo  {journal} {J. Opt. Soc. Am. B}\ }\textbf {\bibinfo {volume}
  {31}},\ \bibinfo {pages} {2460} (\bibinfo {year} {2014})}\BibitemShut
  {NoStop}%
\bibitem [{\citenamefont {Grelu}\ and\ \citenamefont
  {Akhmediev}(2012)}]{Grelu_NP_2012}%
  \BibitemOpen
  \bibfield  {author} {\bibinfo {author} {\bibfnamefont {P.}~\bibnamefont
  {Grelu}}\ and\ \bibinfo {author} {\bibfnamefont {N.}~\bibnamefont
  {Akhmediev}},\ }\href@noop {} {\bibfield  {journal} {\bibinfo  {journal}
  {Nat. Phot.}\ }\textbf {\bibinfo {volume} {6}},\ \bibinfo {pages} {84}
  (\bibinfo {year} {2012})}\BibitemShut {NoStop}%
\bibitem [{\citenamefont {Boardman}\ and\ \citenamefont
  {Xie}(1997)}]{Boardman_1997}%
  \BibitemOpen
  \bibfield  {author} {\bibinfo {author} {\bibfnamefont {A.~D.}\ \bibnamefont
  {Boardman}}\ and\ \bibinfo {author} {\bibfnamefont {K.}~\bibnamefont {Xie}},\
  }\href {\doibase 10.1364/JOSAB.14.003102} {\bibfield  {journal} {\bibinfo
  {journal} {J. Opt. Soc. Am. B}\ }\textbf {\bibinfo {volume} {14}},\ \bibinfo
  {pages} {3102} (\bibinfo {year} {1997})}\BibitemShut {NoStop}%
\bibitem [{\citenamefont {Boardman}\ \emph {et~al.}(2005)\citenamefont
  {Boardman}, \citenamefont {Velasco},\ and\ \citenamefont
  {Egan}}]{Boardman_Chapter_2005}%
  \BibitemOpen
  \bibfield  {author} {\bibinfo {author} {\bibfnamefont {A.}~\bibnamefont
  {Boardman}}, \bibinfo {author} {\bibfnamefont {L.}~\bibnamefont {Velasco}}, \
  and\ \bibinfo {author} {\bibfnamefont {P.}~\bibnamefont {Egan}},\ }\enquote
  {\bibinfo {title} {Dissipative magneto-optic solitons},}\ in\ \href {\doibase
  10.1007/10928028_2} {\emph {\bibinfo {booktitle} {Dissipative Solitons}}},\
  \bibinfo {editor} {edited by\ \bibinfo {editor} {\bibfnamefont
  {N.}~\bibnamefont {Akhmediev}}\ and\ \bibinfo {editor} {\bibfnamefont
  {A.}~\bibnamefont {Ankiewicz}}}\ (\bibinfo  {publisher} {Springer},\ \bibinfo
  {address} {Berlin, Heidelberg},\ \bibinfo {year} {2005})\ pp.\ \bibinfo
  {pages} {19--35}\BibitemShut {NoStop}%
\bibitem [{\citenamefont {Boardman}\ and\ \citenamefont
  {Velasco}(2006)}]{Boardman_2006}%
  \BibitemOpen
  \bibfield  {author} {\bibinfo {author} {\bibfnamefont {A.~D.}\ \bibnamefont
  {Boardman}}\ and\ \bibinfo {author} {\bibfnamefont {L.}~\bibnamefont
  {Velasco}},\ }\href {\doibase 10.1109/JSTQE.2006.872718} {\bibfield
  {journal} {\bibinfo  {journal} {IEEE J. Sel. Top. Quantum Electron.}\
  }\textbf {\bibinfo {volume} {12}},\ \bibinfo {pages} {388} (\bibinfo {year}
  {2006})}\BibitemShut {NoStop}%
\bibitem [{\citenamefont {Liehr}(2013)}]{Liehr_Book}%
  \BibitemOpen
  \bibfield  {author} {\bibinfo {author} {\bibfnamefont {A.}~\bibnamefont
  {Liehr}},\ }\href@noop {} {\emph {\bibinfo {title} {Dissipative Solitons in
  Reaction Diffusion Systems}}}\ (\bibinfo  {publisher} {Springer},\ \bibinfo
  {address} {Berlin},\ \bibinfo {year} {2013})\BibitemShut {NoStop}%
\bibitem [{\citenamefont {Malomed}(2006)}]{Malomed_Book}%
  \BibitemOpen
  \bibfield  {author} {\bibinfo {author} {\bibfnamefont {B.~A.}\ \bibnamefont
  {Malomed}},\ }\href@noop {} {\emph {\bibinfo {title} {Soliton Management in
  Periodic Systems}}}\ (\bibinfo  {publisher} {Springer},\ \bibinfo {address}
  {Berlin},\ \bibinfo {year} {2006})\BibitemShut {NoStop}%
\bibitem [{\citenamefont {Ostrovskaya}\ \emph {et~al.}(2012)\citenamefont
  {Ostrovskaya}, \citenamefont {Abdullaev}, \citenamefont {Desyatnikov},
  \citenamefont {Fraser},\ and\ \citenamefont
  {Kivshar}}]{Ostrovskaya_PRA_2012}%
  \BibitemOpen
  \bibfield  {author} {\bibinfo {author} {\bibfnamefont {E.~A.}\ \bibnamefont
  {Ostrovskaya}}, \bibinfo {author} {\bibfnamefont {J.}~\bibnamefont
  {Abdullaev}}, \bibinfo {author} {\bibfnamefont {A.~S.}\ \bibnamefont
  {Desyatnikov}}, \bibinfo {author} {\bibfnamefont {M.~D.}\ \bibnamefont
  {Fraser}}, \ and\ \bibinfo {author} {\bibfnamefont {Y.~S.}\ \bibnamefont
  {Kivshar}},\ }\href {\doibase 10.1103/PhysRevA.86.013636} {\bibfield
  {journal} {\bibinfo  {journal} {Phys. Rev. A}\ }\textbf {\bibinfo {volume}
  {86}},\ \bibinfo {pages} {013636} (\bibinfo {year} {2012})}\BibitemShut
  {NoStop}%
\bibitem [{\citenamefont {Smirnov}\ \emph {et~al.}(2014)\citenamefont
  {Smirnov}, \citenamefont {Smirnova}, \citenamefont {Ostrovskaya},\ and\
  \citenamefont {Kivshar}}]{Smirnov_PRB_2014}%
  \BibitemOpen
  \bibfield  {author} {\bibinfo {author} {\bibfnamefont {L.~A.}\ \bibnamefont
  {Smirnov}}, \bibinfo {author} {\bibfnamefont {D.~A.}\ \bibnamefont
  {Smirnova}}, \bibinfo {author} {\bibfnamefont {E.~A.}\ \bibnamefont
  {Ostrovskaya}}, \ and\ \bibinfo {author} {\bibfnamefont {Y.~S.}\ \bibnamefont
  {Kivshar}},\ }\href {\doibase 10.1103/PhysRevB.89.235310} {\bibfield
  {journal} {\bibinfo  {journal} {Phys. Rev. B}\ }\textbf {\bibinfo {volume}
  {89}},\ \bibinfo {pages} {235310} (\bibinfo {year} {2014})}\BibitemShut
  {NoStop}%
\bibitem [{\citenamefont {Akhmediev}\ and\ \citenamefont {{A. Ankiewicz
  (Eds.)}}(2005)}]{Akhmediev_Book1}%
  \BibitemOpen
  \bibfield  {author} {\bibinfo {author} {\bibfnamefont {N.}~\bibnamefont
  {Akhmediev}}\ and\ \bibinfo {author} {\bibnamefont {{A. Ankiewicz (Eds.)}}},\
  }\href@noop {} {\emph {\bibinfo {title} {Dissipative Solitons}}}\ (\bibinfo
  {publisher} {Springer},\ \bibinfo {address} {Berlin},\ \bibinfo {year}
  {2005})\BibitemShut {NoStop}%
\bibitem [{\citenamefont {{N. Akhmediev and {A. Ankiewicz
  (Eds.)}}}(2008)}]{Akhmediev_Book2}%
  \BibitemOpen
  \bibfield  {author} {\bibinfo {author} {\bibnamefont {{N. Akhmediev and {A.
  Ankiewicz (Eds.)}}}},\ }\href@noop {} {\emph {\bibinfo {title} {Dissipative
  Solitons: From Optics to Biology and Medicine}}}\ (\bibinfo  {publisher}
  {Springer},\ \bibinfo {address} {Berlin},\ \bibinfo {year}
  {2008})\BibitemShut {NoStop}%
\bibitem [{\citenamefont {Purwins}\ \emph {et~al.}(2010)\citenamefont
  {Purwins}, \citenamefont {Bödeker},\ and\ \citenamefont
  {Sh.Amiranashvili}}]{Purwins_AP_2010}%
  \BibitemOpen
  \bibfield  {author} {\bibinfo {author} {\bibfnamefont {H.-G.}\ \bibnamefont
  {Purwins}}, \bibinfo {author} {\bibfnamefont {H.}~\bibnamefont {Bödeker}}, \
  and\ \bibinfo {author} {\bibnamefont {Sh.Amiranashvili}},\ }\href@noop {}
  {\bibfield  {journal} {\bibinfo  {journal} {Advances in Physics}\ }\textbf
  {\bibinfo {volume} {59}},\ \bibinfo {pages} {485} (\bibinfo {year}
  {2010})}\BibitemShut {NoStop}%
\bibitem [{\citenamefont {Akhmediev}\ and\ \citenamefont
  {Ankiewicz}(2005)}]{Akhmediev_Chapter_2005}%
  \BibitemOpen
  \bibfield  {author} {\bibinfo {author} {\bibfnamefont {N.}~\bibnamefont
  {Akhmediev}}\ and\ \bibinfo {author} {\bibfnamefont {A.}~\bibnamefont
  {Ankiewicz}},\ }\enquote {\bibinfo {title} {Dissipative solitons in the
  complex {Ginzburg-Landau} and {Swift-Hohenberg} equations},}\ in\ \href
  {\doibase 10.1007/10928028_1} {\emph {\bibinfo {booktitle} {Dissipative
  Solitons}}},\ \bibinfo {editor} {edited by\ \bibinfo {editor} {\bibfnamefont
  {N.}~\bibnamefont {Akhmediev}}\ and\ \bibinfo {editor} {\bibfnamefont
  {A.}~\bibnamefont {Ankiewicz}}}\ (\bibinfo  {publisher} {Springer Berlin
  Heidelberg},\ \bibinfo {address} {Berlin, Heidelberg},\ \bibinfo {year}
  {2005})\ pp.\ \bibinfo {pages} {1--17}\BibitemShut {NoStop}%
\bibitem [{\citenamefont {Fauve}\ and\ \citenamefont
  {Thual}(1990)}]{Fauve_PRL_1990}%
  \BibitemOpen
  \bibfield  {author} {\bibinfo {author} {\bibfnamefont {S.}~\bibnamefont
  {Fauve}}\ and\ \bibinfo {author} {\bibfnamefont {O.}~\bibnamefont {Thual}},\
  }\href {\doibase 10.1103/PhysRevLett.64.282} {\bibfield  {journal} {\bibinfo
  {journal} {Phys. Rev. Lett.}\ }\textbf {\bibinfo {volume} {64}},\ \bibinfo
  {pages} {282} (\bibinfo {year} {1990})}\BibitemShut {NoStop}%
\bibitem [{\citenamefont {van Saarloos}\ and\ \citenamefont
  {Hohenberg}(1992)}]{van_Saarloos_PD_1992}%
  \BibitemOpen
  \bibfield  {author} {\bibinfo {author} {\bibfnamefont {W.}~\bibnamefont {van
  Saarloos}}\ and\ \bibinfo {author} {\bibfnamefont {P.~C.}\ \bibnamefont
  {Hohenberg}},\ }\href@noop {} {\bibfield  {journal} {\bibinfo  {journal}
  {Physica D}\ }\textbf {\bibinfo {volume} {56}},\ \bibinfo {pages} {303}
  (\bibinfo {year} {1992})}\BibitemShut {NoStop}%
\bibitem [{\citenamefont {Afanasjev}\ \emph {et~al.}(1996)\citenamefont
  {Afanasjev}, \citenamefont {Akhmediev},\ and\ \citenamefont
  {Soto-Crespo}}]{Afanasjev_PRE_1996}%
  \BibitemOpen
  \bibfield  {author} {\bibinfo {author} {\bibfnamefont {V.~V.}\ \bibnamefont
  {Afanasjev}}, \bibinfo {author} {\bibfnamefont {N.}~\bibnamefont
  {Akhmediev}}, \ and\ \bibinfo {author} {\bibfnamefont {J.~M.}\ \bibnamefont
  {Soto-Crespo}},\ }\href {\doibase 10.1103/PhysRevE.53.1931} {\bibfield
  {journal} {\bibinfo  {journal} {Phys. Rev. E}\ }\textbf {\bibinfo {volume}
  {53}},\ \bibinfo {pages} {1931} (\bibinfo {year} {1996})}\BibitemShut
  {NoStop}%
\bibitem [{\citenamefont {Renninger}\ \emph {et~al.}(2008)\citenamefont
  {Renninger}, \citenamefont {Chong},\ and\ \citenamefont
  {Wise}}]{Renninger_PRA_2008}%
  \BibitemOpen
  \bibfield  {author} {\bibinfo {author} {\bibfnamefont {W.~H.}\ \bibnamefont
  {Renninger}}, \bibinfo {author} {\bibfnamefont {A.}~\bibnamefont {Chong}}, \
  and\ \bibinfo {author} {\bibfnamefont {F.~W.}\ \bibnamefont {Wise}},\ }\href
  {\doibase 10.1103/PhysRevA.77.023814} {\bibfield  {journal} {\bibinfo
  {journal} {Phys. Rev. A}\ }\textbf {\bibinfo {volume} {77}},\ \bibinfo
  {pages} {023814} (\bibinfo {year} {2008})}\BibitemShut {NoStop}%
\bibitem [{\citenamefont {Deissler}\ and\ \citenamefont
  {Brand}(1994)}]{Deissler_PRL_1994}%
  \BibitemOpen
  \bibfield  {author} {\bibinfo {author} {\bibfnamefont {R.~J.}\ \bibnamefont
  {Deissler}}\ and\ \bibinfo {author} {\bibfnamefont {H.~R.}\ \bibnamefont
  {Brand}},\ }\href {\doibase 10.1103/PhysRevLett.72.478} {\bibfield  {journal}
  {\bibinfo  {journal} {Phys. Rev. Lett.}\ }\textbf {\bibinfo {volume} {72}},\
  \bibinfo {pages} {478} (\bibinfo {year} {1994})}\BibitemShut {NoStop}%
\bibitem [{\citenamefont {Soto-Crespo}\ \emph {et~al.}(2000)\citenamefont
  {Soto-Crespo}, \citenamefont {Akhmediev},\ and\ \citenamefont
  {Ankiewicz}}]{Soto-Crespo_PRL_2000}%
  \BibitemOpen
  \bibfield  {author} {\bibinfo {author} {\bibfnamefont {J.~M.}\ \bibnamefont
  {Soto-Crespo}}, \bibinfo {author} {\bibfnamefont {N.}~\bibnamefont
  {Akhmediev}}, \ and\ \bibinfo {author} {\bibfnamefont {A.}~\bibnamefont
  {Ankiewicz}},\ }\href {\doibase 10.1103/PhysRevLett.85.2937} {\bibfield
  {journal} {\bibinfo  {journal} {Phys. Rev. Lett.}\ }\textbf {\bibinfo
  {volume} {85}},\ \bibinfo {pages} {2937} (\bibinfo {year}
  {2000})}\BibitemShut {NoStop}%
\bibitem [{\citenamefont {Akhmediev}\ \emph {et~al.}(2001)\citenamefont
  {Akhmediev}, \citenamefont {Soto-Crespo},\ and\ \citenamefont
  {Town}}]{Akhmediev_PRE_2001}%
  \BibitemOpen
  \bibfield  {author} {\bibinfo {author} {\bibfnamefont {N.}~\bibnamefont
  {Akhmediev}}, \bibinfo {author} {\bibfnamefont {J.~M.}\ \bibnamefont
  {Soto-Crespo}}, \ and\ \bibinfo {author} {\bibfnamefont {G.}~\bibnamefont
  {Town}},\ }\href {\doibase 10.1103/PhysRevE.63.056602} {\bibfield  {journal}
  {\bibinfo  {journal} {Phys. Rev. E}\ }\textbf {\bibinfo {volume} {63}},\
  \bibinfo {pages} {056602} (\bibinfo {year} {2001})}\BibitemShut {NoStop}%
\bibitem [{\citenamefont {Soto-Crespo}\ \emph {et~al.}(2001)\citenamefont
  {Soto-Crespo}, \citenamefont {Akhmediev},\ and\ \citenamefont
  {Chiang}}]{Soto-Crespo_PLA_2001}%
  \BibitemOpen
  \bibfield  {author} {\bibinfo {author} {\bibfnamefont {J.~M.}\ \bibnamefont
  {Soto-Crespo}}, \bibinfo {author} {\bibfnamefont {N.}~\bibnamefont
  {Akhmediev}}, \ and\ \bibinfo {author} {\bibfnamefont {K.~S.}\ \bibnamefont
  {Chiang}},\ }\href@noop {} {\bibfield  {journal} {\bibinfo  {journal} {Phys.
  Lett. A}\ }\textbf {\bibinfo {volume} {291}},\ \bibinfo {pages} {115}
  (\bibinfo {year} {2001})}\BibitemShut {NoStop}%
\bibitem [{\citenamefont {Cundiff}\ \emph {et~al.}(2002)\citenamefont
  {Cundiff}, \citenamefont {Soto-Crespo},\ and\ \citenamefont
  {Akhmediev}}]{Cundiff_PRL_2002}%
  \BibitemOpen
  \bibfield  {author} {\bibinfo {author} {\bibfnamefont {S.~T.}\ \bibnamefont
  {Cundiff}}, \bibinfo {author} {\bibfnamefont {J.~M.}\ \bibnamefont
  {Soto-Crespo}}, \ and\ \bibinfo {author} {\bibfnamefont {N.}~\bibnamefont
  {Akhmediev}},\ }\href {\doibase 10.1103/PhysRevLett.88.073903} {\bibfield
  {journal} {\bibinfo  {journal} {Phys. Rev. Lett.}\ }\textbf {\bibinfo
  {volume} {88}},\ \bibinfo {pages} {073903} (\bibinfo {year}
  {2002})}\BibitemShut {NoStop}%
\bibitem [{\citenamefont {Descalzi}\ \emph {et~al.}(2011)\citenamefont
  {Descalzi}, \citenamefont {Cartes}, \citenamefont {Cisternas},\ and\
  \citenamefont {Brand}}]{Descalzi_PRE_2011}%
  \BibitemOpen
  \bibfield  {author} {\bibinfo {author} {\bibfnamefont {O.}~\bibnamefont
  {Descalzi}}, \bibinfo {author} {\bibfnamefont {C.}~\bibnamefont {Cartes}},
  \bibinfo {author} {\bibfnamefont {J.}~\bibnamefont {Cisternas}}, \ and\
  \bibinfo {author} {\bibfnamefont {H.~R.}\ \bibnamefont {Brand}},\ }\href
  {\doibase 10.1103/PhysRevE.83.056214} {\bibfield  {journal} {\bibinfo
  {journal} {Phys. Rev. E}\ }\textbf {\bibinfo {volume} {83}},\ \bibinfo
  {pages} {056214} (\bibinfo {year} {2011})}\BibitemShut {NoStop}%
\bibitem [{\citenamefont {Chang}\ \emph
  {et~al.}(2015{\natexlab{a}})\citenamefont {Chang}, \citenamefont
  {Soto-Crespo}, \citenamefont {Vouzas},\ and\ \citenamefont
  {Akhmediev}}]{Chang_OL_2015}%
  \BibitemOpen
  \bibfield  {author} {\bibinfo {author} {\bibfnamefont {W.}~\bibnamefont
  {Chang}}, \bibinfo {author} {\bibfnamefont {J.~M.}\ \bibnamefont
  {Soto-Crespo}}, \bibinfo {author} {\bibfnamefont {P.}~\bibnamefont {Vouzas}},
  \ and\ \bibinfo {author} {\bibfnamefont {N.}~\bibnamefont {Akhmediev}},\
  }\href {\doibase 10.1364/OL.40.002949} {\bibfield  {journal} {\bibinfo
  {journal} {Opt. Lett.}\ }\textbf {\bibinfo {volume} {40}},\ \bibinfo {pages}
  {2949} (\bibinfo {year} {2015}{\natexlab{a}})}\BibitemShut {NoStop}%
\bibitem [{\citenamefont {Chang}\ \emph
  {et~al.}(2015{\natexlab{b}})\citenamefont {Chang}, \citenamefont
  {Soto-Crespo}, \citenamefont {Vouzas},\ and\ \citenamefont
  {Akhmediev}}]{Chang_JOSAB_2015}%
  \BibitemOpen
  \bibfield  {author} {\bibinfo {author} {\bibfnamefont {W.}~\bibnamefont
  {Chang}}, \bibinfo {author} {\bibfnamefont {J.~M.}\ \bibnamefont
  {Soto-Crespo}}, \bibinfo {author} {\bibfnamefont {P.}~\bibnamefont {Vouzas}},
  \ and\ \bibinfo {author} {\bibfnamefont {N.}~\bibnamefont {Akhmediev}},\
  }\href {\doibase 10.1364/JOSAB.32.001377} {\bibfield  {journal} {\bibinfo
  {journal} {J. Opt. Soc. Am. B}\ }\textbf {\bibinfo {volume} {32}},\ \bibinfo
  {pages} {1377} (\bibinfo {year} {2015}{\natexlab{b}})}\BibitemShut {NoStop}%
\bibitem [{\citenamefont {Soto-Crespo}\ \emph {et~al.}(2017)\citenamefont
  {Soto-Crespo}, \citenamefont {Devine},\ and\ \citenamefont
  {Akhmediev}}]{Soto-Crespo_JOSAB_2017}%
  \BibitemOpen
  \bibfield  {author} {\bibinfo {author} {\bibfnamefont {J.~M.}\ \bibnamefont
  {Soto-Crespo}}, \bibinfo {author} {\bibfnamefont {N.}~\bibnamefont {Devine}},
  \ and\ \bibinfo {author} {\bibfnamefont {N.}~\bibnamefont {Akhmediev}},\
  }\href {\doibase 10.1364/JOSAB.34.001542} {\bibfield  {journal} {\bibinfo
  {journal} {J. Opt. Soc. Am. B}\ }\textbf {\bibinfo {volume} {34}},\ \bibinfo
  {pages} {1542} (\bibinfo {year} {2017})}\BibitemShut {NoStop}%
\bibitem [{\citenamefont {Akhmediev}\ \emph {et~al.}(1997)\citenamefont
  {Akhmediev}, \citenamefont {Ankiewicz},\ and\ \citenamefont
  {Soto-Crespo}}]{Akhmediev_PRL_1997}%
  \BibitemOpen
  \bibfield  {author} {\bibinfo {author} {\bibfnamefont {N.~N.}\ \bibnamefont
  {Akhmediev}}, \bibinfo {author} {\bibfnamefont {A.}~\bibnamefont
  {Ankiewicz}}, \ and\ \bibinfo {author} {\bibfnamefont {J.~M.}\ \bibnamefont
  {Soto-Crespo}},\ }\href {\doibase 10.1103/PhysRevLett.79.4047} {\bibfield
  {journal} {\bibinfo  {journal} {Phys. Rev. Lett.}\ }\textbf {\bibinfo
  {volume} {79}},\ \bibinfo {pages} {4047} (\bibinfo {year}
  {1997})}\BibitemShut {NoStop}%
\bibitem [{\citenamefont {Turaev}\ \emph {et~al.}(2007)\citenamefont {Turaev},
  \citenamefont {Vladimirov},\ and\ \citenamefont {Zelik}}]{Turaev_PRE_2007}%
  \BibitemOpen
  \bibfield  {author} {\bibinfo {author} {\bibfnamefont {D.}~\bibnamefont
  {Turaev}}, \bibinfo {author} {\bibfnamefont {A.~G.}\ \bibnamefont
  {Vladimirov}}, \ and\ \bibinfo {author} {\bibfnamefont {S.}~\bibnamefont
  {Zelik}},\ }\href {\doibase 10.1103/PhysRevE.75.045601} {\bibfield  {journal}
  {\bibinfo  {journal} {Phys. Rev. E}\ }\textbf {\bibinfo {volume} {75}},\
  \bibinfo {pages} {045601} (\bibinfo {year} {2007})}\BibitemShut {NoStop}%
\bibitem [{\citenamefont {Soto-Crespo}\ and\ \citenamefont
  {Akhmediev}(2002)}]{Soto-Crespo_PRE_2002}%
  \BibitemOpen
  \bibfield  {author} {\bibinfo {author} {\bibfnamefont {J.~M.}\ \bibnamefont
  {Soto-Crespo}}\ and\ \bibinfo {author} {\bibfnamefont {N.}~\bibnamefont
  {Akhmediev}},\ }\href {\doibase 10.1103/PhysRevE.66.066610} {\bibfield
  {journal} {\bibinfo  {journal} {Phys. Rev. E}\ }\textbf {\bibinfo {volume}
  {66}},\ \bibinfo {pages} {066610} (\bibinfo {year} {2002})}\BibitemShut
  {NoStop}%
\bibitem [{\citenamefont {Achilleos}\ \emph {et~al.}(2016)\citenamefont
  {Achilleos}, \citenamefont {Bishop}, \citenamefont {Diamantidis},
  \citenamefont {Frantzeskakis}, \citenamefont {Horikis}, \citenamefont
  {Karachalios},\ and\ \citenamefont {Kevrekidis}}]{Achilleos_PRE_2016}%
  \BibitemOpen
  \bibfield  {author} {\bibinfo {author} {\bibfnamefont {V.}~\bibnamefont
  {Achilleos}}, \bibinfo {author} {\bibfnamefont {A.~R.}\ \bibnamefont
  {Bishop}}, \bibinfo {author} {\bibfnamefont {S.}~\bibnamefont {Diamantidis}},
  \bibinfo {author} {\bibfnamefont {D.~J.}\ \bibnamefont {Frantzeskakis}},
  \bibinfo {author} {\bibfnamefont {T.~P.}\ \bibnamefont {Horikis}}, \bibinfo
  {author} {\bibfnamefont {N.~I.}\ \bibnamefont {Karachalios}}, \ and\ \bibinfo
  {author} {\bibfnamefont {P.~G.}\ \bibnamefont {Kevrekidis}},\ }\href
  {\doibase 10.1103/PhysRevE.94.012210} {\bibfield  {journal} {\bibinfo
  {journal} {Phys. Rev. E}\ }\textbf {\bibinfo {volume} {94}},\ \bibinfo
  {pages} {012210} (\bibinfo {year} {2016})}\BibitemShut {NoStop}%
\bibitem [{\citenamefont {Sakaguchi}\ \emph {et~al.}(2018)\citenamefont
  {Sakaguchi}, \citenamefont {Skryabin},\ and\ \citenamefont
  {Malomed}}]{Sakaguchi_OL_2018}%
  \BibitemOpen
  \bibfield  {author} {\bibinfo {author} {\bibfnamefont {H.}~\bibnamefont
  {Sakaguchi}}, \bibinfo {author} {\bibfnamefont {D.~V.}\ \bibnamefont
  {Skryabin}}, \ and\ \bibinfo {author} {\bibfnamefont {B.~A.}\ \bibnamefont
  {Malomed}},\ }\href {\doibase 10.1364/OL.43.002688} {\bibfield  {journal}
  {\bibinfo  {journal} {Opt. Lett.}\ }\textbf {\bibinfo {volume} {43}},\
  \bibinfo {pages} {2688} (\bibinfo {year} {2018})}\BibitemShut {NoStop}%
\bibitem [{\citenamefont {Uzunov}\ \emph {et~al.}(2018)\citenamefont {Uzunov},
  \citenamefont {Georgiev},\ and\ \citenamefont
  {Arabadzhiev}}]{Uzunov_PRE_2018}%
  \BibitemOpen
  \bibfield  {author} {\bibinfo {author} {\bibfnamefont {I.~M.}\ \bibnamefont
  {Uzunov}}, \bibinfo {author} {\bibfnamefont {Z.~D.}\ \bibnamefont
  {Georgiev}}, \ and\ \bibinfo {author} {\bibfnamefont {T.~N.}\ \bibnamefont
  {Arabadzhiev}},\ }\href {\doibase 10.1103/PhysRevE.97.052215} {\bibfield
  {journal} {\bibinfo  {journal} {Phys. Rev. E}\ }\textbf {\bibinfo {volume}
  {97}},\ \bibinfo {pages} {052215} (\bibinfo {year} {2018})}\BibitemShut
  {NoStop}%
\bibitem [{\citenamefont {Crasovan}\ \emph {et~al.}(2000)\citenamefont
  {Crasovan}, \citenamefont {Malomed},\ and\ \citenamefont
  {Mihalache}}]{Crasovan_PRE_2000}%
  \BibitemOpen
  \bibfield  {author} {\bibinfo {author} {\bibfnamefont {L.-C.}\ \bibnamefont
  {Crasovan}}, \bibinfo {author} {\bibfnamefont {B.~A.}\ \bibnamefont
  {Malomed}}, \ and\ \bibinfo {author} {\bibfnamefont {D.}~\bibnamefont
  {Mihalache}},\ }\href {\doibase 10.1103/PhysRevE.63.016605} {\bibfield
  {journal} {\bibinfo  {journal} {Phys. Rev. E}\ }\textbf {\bibinfo {volume}
  {63}},\ \bibinfo {pages} {016605} (\bibinfo {year} {2000})}\BibitemShut
  {NoStop}%
\bibitem [{\citenamefont {Mihalache}\ \emph {et~al.}(2006)\citenamefont
  {Mihalache}, \citenamefont {Mazilu}, \citenamefont {Lederer}, \citenamefont
  {Kartashov}, \citenamefont {Crasovan}, \citenamefont {Torner},\ and\
  \citenamefont {Malomed}}]{Mihalache_PRL_2006}%
  \BibitemOpen
  \bibfield  {author} {\bibinfo {author} {\bibfnamefont {D.}~\bibnamefont
  {Mihalache}}, \bibinfo {author} {\bibfnamefont {D.}~\bibnamefont {Mazilu}},
  \bibinfo {author} {\bibfnamefont {F.}~\bibnamefont {Lederer}}, \bibinfo
  {author} {\bibfnamefont {Y.~V.}\ \bibnamefont {Kartashov}}, \bibinfo {author}
  {\bibfnamefont {L.-C.}\ \bibnamefont {Crasovan}}, \bibinfo {author}
  {\bibfnamefont {L.}~\bibnamefont {Torner}}, \ and\ \bibinfo {author}
  {\bibfnamefont {B.~A.}\ \bibnamefont {Malomed}},\ }\href {\doibase
  10.1103/PhysRevLett.97.073904} {\bibfield  {journal} {\bibinfo  {journal}
  {Phys. Rev. Lett.}\ }\textbf {\bibinfo {volume} {97}},\ \bibinfo {pages}
  {073904} (\bibinfo {year} {2006})}\BibitemShut {NoStop}%
\bibitem [{\citenamefont {Soto-Crespo}\ \emph {et~al.}(2006)\citenamefont
  {Soto-Crespo}, \citenamefont {Grelu},\ and\ \citenamefont
  {Akhmediev}}]{Soto-Crespo_OE_2006}%
  \BibitemOpen
  \bibfield  {author} {\bibinfo {author} {\bibfnamefont {J.~M.}\ \bibnamefont
  {Soto-Crespo}}, \bibinfo {author} {\bibfnamefont {P.}~\bibnamefont {Grelu}},
  \ and\ \bibinfo {author} {\bibfnamefont {N.}~\bibnamefont {Akhmediev}},\
  }\href {\doibase 10.1364/OE.14.004013} {\bibfield  {journal} {\bibinfo
  {journal} {Opt. Express}\ }\textbf {\bibinfo {volume} {14}},\ \bibinfo
  {pages} {4013} (\bibinfo {year} {2006})}\BibitemShut {NoStop}%
\bibitem [{\citenamefont {Akhmediev}\ \emph {et~al.}(2007)\citenamefont
  {Akhmediev}, \citenamefont {Soto-Crespo},\ and\ \citenamefont
  {Grelu}}]{Akhmediev_Chaos_2007}%
  \BibitemOpen
  \bibfield  {author} {\bibinfo {author} {\bibfnamefont {N.}~\bibnamefont
  {Akhmediev}}, \bibinfo {author} {\bibfnamefont {J.~M.}\ \bibnamefont
  {Soto-Crespo}}, \ and\ \bibinfo {author} {\bibfnamefont {P.}~\bibnamefont
  {Grelu}},\ }\href@noop {} {\bibfield  {journal} {\bibinfo  {journal} {Chaos}\
  }\textbf {\bibinfo {volume} {17}},\ \bibinfo {pages} {037112} (\bibinfo
  {year} {2007})}\BibitemShut {NoStop}%
\bibitem [{\citenamefont {Mihalache}\ \emph {et~al.}(2007)\citenamefont
  {Mihalache}, \citenamefont {Mazilu}, \citenamefont {Lederer}, \citenamefont
  {Leblond},\ and\ \citenamefont {Malomed}}]{Mihalache_PRA_2007}%
  \BibitemOpen
  \bibfield  {author} {\bibinfo {author} {\bibfnamefont {D.}~\bibnamefont
  {Mihalache}}, \bibinfo {author} {\bibfnamefont {D.}~\bibnamefont {Mazilu}},
  \bibinfo {author} {\bibfnamefont {F.}~\bibnamefont {Lederer}}, \bibinfo
  {author} {\bibfnamefont {H.}~\bibnamefont {Leblond}}, \ and\ \bibinfo
  {author} {\bibfnamefont {B.~A.}\ \bibnamefont {Malomed}},\ }\href {\doibase
  10.1103/PhysRevA.75.033811} {\bibfield  {journal} {\bibinfo  {journal} {Phys.
  Rev. A}\ }\textbf {\bibinfo {volume} {75}},\ \bibinfo {pages} {033811}
  (\bibinfo {year} {2007})}\BibitemShut {NoStop}%
\bibitem [{\citenamefont {Soskin}\ \emph {et~al.}(2017)\citenamefont {Soskin},
  \citenamefont {Boriskina}, \citenamefont {Chong}, \citenamefont {Dennis},\
  and\ \citenamefont {Desyatnikov}}]{Soskin_JO_2017}%
  \BibitemOpen
  \bibfield  {author} {\bibinfo {author} {\bibfnamefont {M.}~\bibnamefont
  {Soskin}}, \bibinfo {author} {\bibfnamefont {S.~V.}\ \bibnamefont
  {Boriskina}}, \bibinfo {author} {\bibfnamefont {Y.}~\bibnamefont {Chong}},
  \bibinfo {author} {\bibfnamefont {M.~R.}\ \bibnamefont {Dennis}}, \ and\
  \bibinfo {author} {\bibfnamefont {A.}~\bibnamefont {Desyatnikov}},\ }\href
  {http://stacks.iop.org/2040-8986/19/i=1/a=010401} {\bibfield  {journal}
  {\bibinfo  {journal} {Journal of Optics}\ }\textbf {\bibinfo {volume} {19}},\
  \bibinfo {pages} {010401} (\bibinfo {year} {2017})}\BibitemShut {NoStop}%
\bibitem [{\citenamefont {Leblond}\ \emph {et~al.}(2009)\citenamefont
  {Leblond}, \citenamefont {Malomed},\ and\ \citenamefont
  {Mihalache}}]{Leblond_PRA_2009}%
  \BibitemOpen
  \bibfield  {author} {\bibinfo {author} {\bibfnamefont {H.}~\bibnamefont
  {Leblond}}, \bibinfo {author} {\bibfnamefont {B.~A.}\ \bibnamefont
  {Malomed}}, \ and\ \bibinfo {author} {\bibfnamefont {D.}~\bibnamefont
  {Mihalache}},\ }\href {\doibase 10.1103/PhysRevA.80.033835} {\bibfield
  {journal} {\bibinfo  {journal} {Phys. Rev. A}\ }\textbf {\bibinfo {volume}
  {80}},\ \bibinfo {pages} {033835} (\bibinfo {year} {2009})}\BibitemShut
  {NoStop}%
\bibitem [{\citenamefont {Sakaguchi}\ and\ \citenamefont
  {Malomed}(2009)}]{Sakaguchi_PRE_2009}%
  \BibitemOpen
  \bibfield  {author} {\bibinfo {author} {\bibfnamefont {H.}~\bibnamefont
  {Sakaguchi}}\ and\ \bibinfo {author} {\bibfnamefont {B.~A.}\ \bibnamefont
  {Malomed}},\ }\href {\doibase 10.1103/PhysRevE.80.026606} {\bibfield
  {journal} {\bibinfo  {journal} {Phys. Rev. E}\ }\textbf {\bibinfo {volume}
  {80}},\ \bibinfo {pages} {026606} (\bibinfo {year} {2009})}\BibitemShut
  {NoStop}%
\bibitem [{\citenamefont {He}\ \emph {et~al.}(2009)\citenamefont {He},
  \citenamefont {Malomed}, \citenamefont {Mihalache}, \citenamefont {Liu},
  \citenamefont {Huang}, \citenamefont {Yang},\ and\ \citenamefont
  {Wang}}]{He_OL_2009}%
  \BibitemOpen
  \bibfield  {author} {\bibinfo {author} {\bibfnamefont {Y.~J.}\ \bibnamefont
  {He}}, \bibinfo {author} {\bibfnamefont {B.~A.}\ \bibnamefont {Malomed}},
  \bibinfo {author} {\bibfnamefont {D.}~\bibnamefont {Mihalache}}, \bibinfo
  {author} {\bibfnamefont {B.}~\bibnamefont {Liu}}, \bibinfo {author}
  {\bibfnamefont {H.~C.}\ \bibnamefont {Huang}}, \bibinfo {author}
  {\bibfnamefont {H.}~\bibnamefont {Yang}}, \ and\ \bibinfo {author}
  {\bibfnamefont {H.~Z.}\ \bibnamefont {Wang}},\ }\href {\doibase
  10.1364/OL.34.002976} {\bibfield  {journal} {\bibinfo  {journal} {Opt.
  Lett.}\ }\textbf {\bibinfo {volume} {34}},\ \bibinfo {pages} {2976} (\bibinfo
  {year} {2009})}\BibitemShut {NoStop}%
\bibitem [{\citenamefont {He}\ \emph {et~al.}(2010)\citenamefont {He},
  \citenamefont {Malomed}, \citenamefont {Ye},\ and\ \citenamefont
  {Hu}}]{He_JOSAB_2010}%
  \BibitemOpen
  \bibfield  {author} {\bibinfo {author} {\bibfnamefont {Y.-J.}\ \bibnamefont
  {He}}, \bibinfo {author} {\bibfnamefont {B.~A.}\ \bibnamefont {Malomed}},
  \bibinfo {author} {\bibfnamefont {F.}~\bibnamefont {Ye}}, \ and\ \bibinfo
  {author} {\bibfnamefont {B.}~\bibnamefont {Hu}},\ }\href {\doibase
  10.1364/JOSAB.27.001139} {\bibfield  {journal} {\bibinfo  {journal} {J. Opt.
  Soc. Am. B}\ }\textbf {\bibinfo {volume} {27}},\ \bibinfo {pages} {1139}
  (\bibinfo {year} {2010})}\BibitemShut {NoStop}%
\bibitem [{\citenamefont {Liu}\ \emph {et~al.}(2010)\citenamefont {Liu},
  \citenamefont {He}, \citenamefont {Malomed}, \citenamefont {Wang},
  \citenamefont {Kevrekidis}, \citenamefont {Wang}, \citenamefont {Leng},
  \citenamefont {Qiu},\ and\ \citenamefont {Wang}}]{Liu_OL_2010}%
  \BibitemOpen
  \bibfield  {author} {\bibinfo {author} {\bibfnamefont {B.}~\bibnamefont
  {Liu}}, \bibinfo {author} {\bibfnamefont {Y.-J.}\ \bibnamefont {He}},
  \bibinfo {author} {\bibfnamefont {B.~A.}\ \bibnamefont {Malomed}}, \bibinfo
  {author} {\bibfnamefont {X.-S.}\ \bibnamefont {Wang}}, \bibinfo {author}
  {\bibfnamefont {P.~G.}\ \bibnamefont {Kevrekidis}}, \bibinfo {author}
  {\bibfnamefont {T.-B.}\ \bibnamefont {Wang}}, \bibinfo {author}
  {\bibfnamefont {F.-C.}\ \bibnamefont {Leng}}, \bibinfo {author}
  {\bibfnamefont {Z.-R.}\ \bibnamefont {Qiu}}, \ and\ \bibinfo {author}
  {\bibfnamefont {H.-Z.}\ \bibnamefont {Wang}},\ }\href {\doibase
  10.1364/OL.35.001974} {\bibfield  {journal} {\bibinfo  {journal} {Opt.
  Lett.}\ }\textbf {\bibinfo {volume} {35}},\ \bibinfo {pages} {1974} (\bibinfo
  {year} {2010})}\BibitemShut {NoStop}%
\bibitem [{\citenamefont {Yin}\ \emph {et~al.}(2011)\citenamefont {Yin},
  \citenamefont {Mihalache},\ and\ \citenamefont {He}}]{Yin_JOSAB_2011}%
  \BibitemOpen
  \bibfield  {author} {\bibinfo {author} {\bibfnamefont {C.}~\bibnamefont
  {Yin}}, \bibinfo {author} {\bibfnamefont {D.}~\bibnamefont {Mihalache}}, \
  and\ \bibinfo {author} {\bibfnamefont {Y.}~\bibnamefont {He}},\ }\href
  {\doibase 10.1364/JOSAB.28.000342} {\bibfield  {journal} {\bibinfo  {journal}
  {J. Opt. Soc. Am. B}\ }\textbf {\bibinfo {volume} {28}},\ \bibinfo {pages}
  {342} (\bibinfo {year} {2011})}\BibitemShut {NoStop}%
\bibitem [{\citenamefont {Liu}\ \emph {et~al.}(2013)\citenamefont {Liu},
  \citenamefont {He},\ and\ \citenamefont {Li}}]{Liu_OE_2013}%
  \BibitemOpen
  \bibfield  {author} {\bibinfo {author} {\bibfnamefont {B.}~\bibnamefont
  {Liu}}, \bibinfo {author} {\bibfnamefont {X.-D.}\ \bibnamefont {He}}, \ and\
  \bibinfo {author} {\bibfnamefont {S.-J.}\ \bibnamefont {Li}},\ }\href
  {\doibase 10.1364/OE.21.005561} {\bibfield  {journal} {\bibinfo  {journal}
  {Opt. Express}\ }\textbf {\bibinfo {volume} {21}},\ \bibinfo {pages} {5561}
  (\bibinfo {year} {2013})}\BibitemShut {NoStop}%
\bibitem [{\citenamefont {Boardman}\ and\ \citenamefont
  {Xie}(2001)}]{Boardman_2001}%
  \BibitemOpen
  \bibfield  {author} {\bibinfo {author} {\bibfnamefont {A.~D.}\ \bibnamefont
  {Boardman}}\ and\ \bibinfo {author} {\bibfnamefont {M.}~\bibnamefont {Xie}},\
  }\href {http://stacks.iop.org/1464-4266/3/i=2/a=376} {\bibfield  {journal}
  {\bibinfo  {journal} {J. Opt. B: Quantum Semiclass. Opt.}\ }\textbf {\bibinfo
  {volume} {3}},\ \bibinfo {pages} {S244} (\bibinfo {year} {2001})}\BibitemShut
  {NoStop}%
\bibitem [{\citenamefont {Kochetov}\ \emph {et~al.}(2017)\citenamefont
  {Kochetov}, \citenamefont {Vasylieva}, \citenamefont {Kochetova},
  \citenamefont {Sun},\ and\ \citenamefont {Tuz}}]{OptLett_2017}%
  \BibitemOpen
  \bibfield  {author} {\bibinfo {author} {\bibfnamefont {B.~A.}\ \bibnamefont
  {Kochetov}}, \bibinfo {author} {\bibfnamefont {I.}~\bibnamefont {Vasylieva}},
  \bibinfo {author} {\bibfnamefont {L.~A.}\ \bibnamefont {Kochetova}}, \bibinfo
  {author} {\bibfnamefont {H.-B.}\ \bibnamefont {Sun}}, \ and\ \bibinfo
  {author} {\bibfnamefont {V.~R.}\ \bibnamefont {Tuz}},\ }\href {\doibase
  10.1364/OL.42.000531} {\bibfield  {journal} {\bibinfo  {journal} {Opt.
  Lett.}\ }\textbf {\bibinfo {volume} {42}},\ \bibinfo {pages} {531} (\bibinfo
  {year} {2017})}\BibitemShut {NoStop}%
\bibitem [{\citenamefont {Kochetov}\ and\ \citenamefont
  {Tuz}(2017)}]{PRE_2017}%
  \BibitemOpen
  \bibfield  {author} {\bibinfo {author} {\bibfnamefont {B.~A.}\ \bibnamefont
  {Kochetov}}\ and\ \bibinfo {author} {\bibfnamefont {V.~R.}\ \bibnamefont
  {Tuz}},\ }\href {\doibase 10.1103/PhysRevE.96.012206} {\bibfield  {journal}
  {\bibinfo  {journal} {Phys. Rev. E}\ }\textbf {\bibinfo {volume} {96}},\
  \bibinfo {pages} {012206} (\bibinfo {year} {2017})}\BibitemShut {NoStop}%
\bibitem [{\citenamefont {Kochetov}\ and\ \citenamefont
  {Tuz}(2018)}]{Chaos_2018}%
  \BibitemOpen
  \bibfield  {author} {\bibinfo {author} {\bibfnamefont {B.~A.}\ \bibnamefont
  {Kochetov}}\ and\ \bibinfo {author} {\bibfnamefont {V.~R.}\ \bibnamefont
  {Tuz}},\ }\href {\doibase 10.1063/1.5016914} {\bibfield  {journal} {\bibinfo
  {journal} {Chaos}\ }\textbf {\bibinfo {volume} {28}},\ \bibinfo {pages}
  {013130} (\bibinfo {year} {2018})}\BibitemShut {NoStop}%
\bibitem [{\citenamefont {Holmer}\ \emph {et~al.}(2007)\citenamefont {Holmer},
  \citenamefont {Marzuola},\ and\ \citenamefont {Zworski}}]{Holmer_JNS_2007}%
  \BibitemOpen
  \bibfield  {author} {\bibinfo {author} {\bibfnamefont {J.}~\bibnamefont
  {Holmer}}, \bibinfo {author} {\bibfnamefont {J.}~\bibnamefont {Marzuola}}, \
  and\ \bibinfo {author} {\bibfnamefont {M.}~\bibnamefont {Zworski}},\ }\href
  {\doibase 10.1007/s00332-006-0807-9} {\bibfield  {journal} {\bibinfo
  {journal} {J. Nonlinear Sci.}\ }\textbf {\bibinfo {volume} {17}},\ \bibinfo
  {pages} {349} (\bibinfo {year} {2007})}\BibitemShut {NoStop}%
\bibitem [{\citenamefont {Yang}\ and\ \citenamefont {Wu}(2008)}]{Yang_OE_2008}%
  \BibitemOpen
  \bibfield  {author} {\bibinfo {author} {\bibfnamefont {R.}~\bibnamefont
  {Yang}}\ and\ \bibinfo {author} {\bibfnamefont {X.}~\bibnamefont {Wu}},\
  }\href {\doibase 10.1364/OE.16.017759} {\bibfield  {journal} {\bibinfo
  {journal} {Opt. Express}\ }\textbf {\bibinfo {volume} {16}},\ \bibinfo
  {pages} {17759} (\bibinfo {year} {2008})}\BibitemShut {NoStop}%
\bibitem [{\citenamefont {Liu}\ and\ \citenamefont {He}(2011)}]{Liu_OE_2011}%
  \BibitemOpen
  \bibfield  {author} {\bibinfo {author} {\bibfnamefont {B.}~\bibnamefont
  {Liu}}\ and\ \bibinfo {author} {\bibfnamefont {X.-D.}\ \bibnamefont {He}},\
  }\href {\doibase 10.1364/OE.19.020009} {\bibfield  {journal} {\bibinfo
  {journal} {Opt. Express}\ }\textbf {\bibinfo {volume} {19}},\ \bibinfo
  {pages} {20009} (\bibinfo {year} {2011})}\BibitemShut {NoStop}%
\bibitem [{\citenamefont {Cox}\ and\ \citenamefont
  {Matthews}(2002)}]{Cox_2002}%
  \BibitemOpen
  \bibfield  {author} {\bibinfo {author} {\bibfnamefont {S.}~\bibnamefont
  {Cox}}\ and\ \bibinfo {author} {\bibfnamefont {P.}~\bibnamefont {Matthews}},\
  }\href {\doibase http://dx.doi.org/10.1006/jcph.2002.6995} {\bibfield
  {journal} {\bibinfo  {journal} {J. Comput. Phys.}\ }\textbf {\bibinfo
  {volume} {176}},\ \bibinfo {pages} {430} (\bibinfo {year}
  {2002})}\BibitemShut {NoStop}%
\bibitem [{Sup()}]{Suppl_Mat}%
  \BibitemOpen
  \href@noop {} {}\bibinfo {note} {See Supplemental Material at (url), (url),
  and (url) for visualization of replication of fundamental dissipative
  soliton, dissipative vortex, and dissipative antivortex,
  respectively.}\BibitemShut {Stop}%
\end{thebibliography}%

\end{document}